\documentclass{aa}

\usepackage[varg]{txfonts}
\usepackage{booktabs}
\usepackage{threeparttable}
\usepackage{multirow}
\usepackage{siunitx}
\bibpunct{(}{)}{;}{a}{}{,} 
\begin{document}

\title{The [Y/Mg] clock works for evolved solar metallicity stars
\thanks{Based on spectroscopic observations made with two telescopes: the Nordic Optical Telescope operated by NOTSA at the Observatorio del Roque de los Muchachos (La Palma, Spain) of the Instituto de Astrofísica de Canarias and the Keck I Telescope at the W.M. Keck Observatory (Mauna Kea, Hawaii, USA) operated by the California Institute of Technology, the University of California and the National Aeronautics and Space Administration.}
}
\author{D. Slumstrup \inst{1} \and F. Grundahl \inst{1} \and K. Brogaard \inst{1,2} \and A. O. Thygesen \inst{3} \and P. E. Nissen \inst{1} \and J. Jessen-Hansen \inst{1} \and V. Van Eylen \inst{4} \and  \\ M. G. Pedersen \inst{5}}

\institute{Stellar Astrophysics Centre (SAC). Department of Physics and Astronomy, Aarhus University, Ny Munkegade 120, DK-8000 Aarhus, Denmark \email{ditte@phys.au.dk}
\and School of Physics and Astronomy, University of Birmingham, Edgbaston, Birmingham B15 2TT, UK
\and California Institute of Technology, 1200 E. California Blvd, MC 249-17, Pasadena, CA 91125, USA
\and Leiden Observatory, Leiden University, 2333CA Leiden, The Netherlands
\and Instituut voor Sterrenkunde, KU Leuven, Celestijnenlaan 200D, 3001 Leuven, Belgium} 
\date{Received 3 July 2017 / Accepted 20 July2017 }

\abstract{}
{Previously [Y/Mg] has been proven to be an age indicator for solar twins. Here, we investigate if this relation also holds for helium-core-burning stars of solar metallicity.
}
{High resolution and high signal-to-noise ratio (S/N) spectroscopic data of stars in the helium-core-burning phase have been obtained with the FIES spectrograph on the NOT \SI{2.56}{m} telescope and the HIRES spectrograph on the Keck I \SI{10}{m} telescope. They have been analyzed to determine the chemical abundances of four open clusters with close to solar metallicity; NGC\,6811, NGC\,6819, M67 and NGC\,188. The abundances are derived from equivalent widths of spectral lines using ATLAS9 model atmospheres with parameters determined from the excitation and ionization balance of Fe lines. Results from asteroseismology and binary studies were used as priors on the atmospheric parameters, where especially the $\log g$ is determined to much higher precision than what is possible with spectroscopy.
}
{It is confirmed that the four open clusters are close to solar metallicity and they follow the [Y/Mg] vs. age trend previously found for solar twins.
}
{The [Y/Mg] vs. age clock also works for giant stars in the helium-core burning phase, which vastly increases the possibilities to estimate the age of stars not only in the solar neighborhood, but in large parts of the Galaxy, due to the brighter nature of evolved stars compared to dwarfs.
}

\keywords{stars: abundances - stars: fundamental parameters - stars: late-type - Galaxy: evolution - open clusters and associations: individual: NGC\,6811, NGC\,6819, M67, NGC\,188}

\maketitle

\section{Introduction}

Stars in an open cluster can be assumed to originate from the same molecular cloud and are therefore expected to have the same chemical composition and age. Age is an especially difficult parameter to determine for single field stars and different methods yield ages that are not necessarily in agreement \citep{Soderblom2010}, whereas stars in clusters have ages determined to much higher precision (e.g., \citet{Vandenberg2013}).  

\citet{daSilva2012}, \citet{Nissen2016} and \citet{TucciMaia2016} investigated trends of different chemical abundances with stellar age and found that [Y/Mg] is a sensitive age indicator for solar twins.
\citet{Feltzing2016} did an independent study extending the sample to a much larger range in [Fe/H]. They found that the relation depends on [Fe/H]. For solar metallicity stars they clearly see the relation, but for stars with [Fe/H] $\sim$ -0.5 dex the relation is insignificant and cannot be used to determine age. Our targets have solar metallicity but are evolved (in the helium-core-burning phase) and therefore do not have the same atmospheric properties as the solar twins. They are also located at larger distances than previously studied. Ages are well determined from cluster studies, which allows us to determine if the relation found for solar twins persists for helium-core-burning stars. 
Previously, yttrium abundances of open clusters have been studied to find possible trends with age, for example,  \citet{Mishenina2014} found no clear trend of [Y/Fe] with age most likely due to large uncertainties, whereas \citet{Maiorca2011} find a declining trend, but the scatter is large, which could be affected by the large range in [Fe/H].

We have carried out a detailed fundamental parameter and abundance analysis of six targets in four open clusters (NGC\,6811, NGC\,6819, M67 and NGC\,188) based on high-resolution and high signal-to-noise ratio (S/N) spectroscopic data from the Nordic Optical Telescope (NOT) and the Keck I Telescope. For NGC\,6811, NGC\,6819 and M67 we have asteroseismic data available from the \textit{Kepler} \citep{Borucki1997} and K2 missions \citep{Howell2014}, which put a strong constraint on the $\log g$ value and can thereby constrain the analysis.

In Sect.~\ref{sec:obs} we present the data taken for this project. In Sect.~\ref{sec:analysis}, we describe the analysis. In Sect.~\ref{sec:results} we present the results. In Sect.~\ref{sec:ymgage}, we test the [Y/Mg] vs. age relation by \citet{Nissen2016} for our sample of solar metallicity helium-core-burning stars. Finally, we conclude on the results in Sect.~\ref{sec:conclusion}.

\section{Targets, observations and data reduction}
\label{sec:obs}

The metallicity of NGC\,6819 is still debated \citep{Bragaglia2001,Lee-Brown2015} and we aim to establish it securely through the analysis in this project. M67 is a very well studied nearby solar-like cluster, which has also been observed with the K2 mission. NGC\,188 is also a solar metallicity cluster that is older and fairly well studied, however not as well as M67. For these three clusters, we have only chosen one target in each cluster, but they are all confirmed members, \citep{Hole2009,Yadav2008,Stetson2004}. The data used in our analysis for both NGC\,188 and NGC\,6819 is of higher resolution and higher S/N than previously used. NGC\,6811 is also a solar-metallicity cluster and it is the youngest in the sample. For this we have three targets, all confirmed members \citep{Sandquist2016}. All targets, except the NGC\,188 target, is confirmed by asteroseismology to be in the helium-core-burning phase \citep{Corsaro2012,Stello2016,Arentoft2017}.
We have adopted literature values for reddening, distance modulus and age for each cluster (see Table~\ref{tab:cluster}).
\begin{table}\small
        \centering
        \caption{\textsl{Cluster information.}}
        \begin{threeparttable}
                \begin{tabular}{l c c c}
                        \toprule
                        Cluster & $E(B-V)$ & $(m-M)_V$ & Age [Gyr]\\
                        \midrule
                        NGC\,6811\tnote{a} & 0.05 & 10.3 & 1.0 $\pm$ 0.1\\
                        NGC\,6819\tnote{b} & 0.16 & 12.4 & 2.4 $\pm$ 0.3\\
                        M67\tnote{c}       & 0.04 & 9.7  & 4.1 $\pm$ 0.5\\
                        NGC\,188\tnote{d}  & 0.09 & 11.2 & 6.2 $\pm$ 0.3\\
                        \bottomrule
                \end{tabular}
                \footnotesize
                \begin{tablenotes}
                \item[a] \citet{Molenda-Zakowicz2014}
                \item[b] \citet{Rosvick1998}
                \item[c] Reddening from \citet{Taylor2007}, distance modulus and age from \citet{Yadav2008}
                \item[d] \citet{Meibom2009}
                \end{tablenotes}
        \end{threeparttable}
        \label{tab:cluster}
\end{table}

\begin{table*}\small
        \centering
        \caption{\textsl{Observations information.}}
        \begin{threeparttable}
                \begin{tabular}{l c c c c c c c}
                        \toprule
                        \multirow{2}{*}{Target} & \multirow{2}{*}{$\alpha_{J2000}$} & \multirow{2}{*}{$\delta_{J2000}$} & \multirow{2}{*}{$V$} & \multirow{2}{*}{Memb. prop} & \multirow{2}{1.1cm}{Number\newline of exp.} & \multirow{2}{1.6cm}{Total exp.\newline time [h]} & \multirow{2}{*}{S/N @6150\AA}\\
                        \\
                        \midrule
                        NGC\,6811-KIC9655167   & 19 37 02.68 & +46 23 13.1 & 11.32 & 97 & 1 & 0.31 & 145\\
                        NGC\,6811-KIC9716090   & 19 36 55.80 & +46 27 37.6 & 11.53 & 94 & 1 & 0.51 & 140\\
                        NGC\,6811-KIC9716522   & 19 37 34.63 & +46 24 10.1 & 10.65 & 97 & 1 & 0.15 & 150\\
                        \midrule 
                        NGC\,6819-KIC5024327   & 19 41 13.45 & +40 11 56.2 & 13.11 & 94 & 21  & 17.0 & 115\\
                        M67-EPIC211415732      & 08 51 12.70 & +11 52 42.4 & 10.39 & 97 & 7   & 3.9  & 135\\
                        NGC\,188-5085\tnote{b} & 00 46 59.57 & +85 13 15.8 & 12.31 & 85 & 13  & 11.5 & 105\\
                        \bottomrule
                \end{tabular}
                \footnotesize
                \begin{tablenotes}
                \item[b] \citet{Stetson2004}
                \end{tablenotes}
        \end{threeparttable}
        \label{tab:obs}
\end{table*}

The observations of NGC\,6819, M67 and NGC\,188 were collected with the Nordic Optical Telescope (NOT) on La Palma, Spain. Spectra for the three stars were obtained in the summer of 2013 and 2015 with the high resolution FIbre-fed Echelle Spectrograph (FIES) covering the wavelength region from 3700-7300\AA, see \citet{Telting2014} for a detailed description of the spectrograph. All observations were carried out in the high resolution mode, $R=67,000$. 
The individual spectra were reduced with FIEStool\footnote{\scriptsize \url{http://www.not.iac.es/instruments/fies/fiestool/}}, an automated data reduction software for FIES.  
Lastly, the spectra for each target were shifted to a common wavelength scale and merged.

The observations of the three targets in NGC\,6811 were carried out on the night of Aug. 23, 2016 using the HIRES spectrograph \citep{Vogt1994} on Keck I. All targets were observed with the red cross-disperser, covering the wavelength region from 4000-7900\AA, with a few inter-order gaps for the reddest orders. We used the ''C5'' decker, providing a resolution of $R=37\,000$. The targets were observed with the exposure meter in operation \citep{Kibrick2006}  to ensure a uniform S/N in all spectra. 
The reductions were performed using the MAKEE pipeline\footnote{\scriptsize \url{http://www.astro.caltech.edu/~tb/makee/}}. All observations are presented in Table~\ref{tab:obs}.

The co-added spectrum for each star was normalized order by order using RAINBOW\footnote{\scriptsize \url{http://sites.google.com/site/vikingpowersoftware}}, which uses appropriate synthetic spectra to identify continuum points in the observed spectrum, which are then fitted with a spline function. The S/N values given in Table~\ref{tab:obs} were estimated from the rms variation of the flux in a region around 6150 Å.

\section{Data analysis}
\label{sec:analysis}

\begin{figure*}
\centering
\includegraphics[width=.24\textwidth]{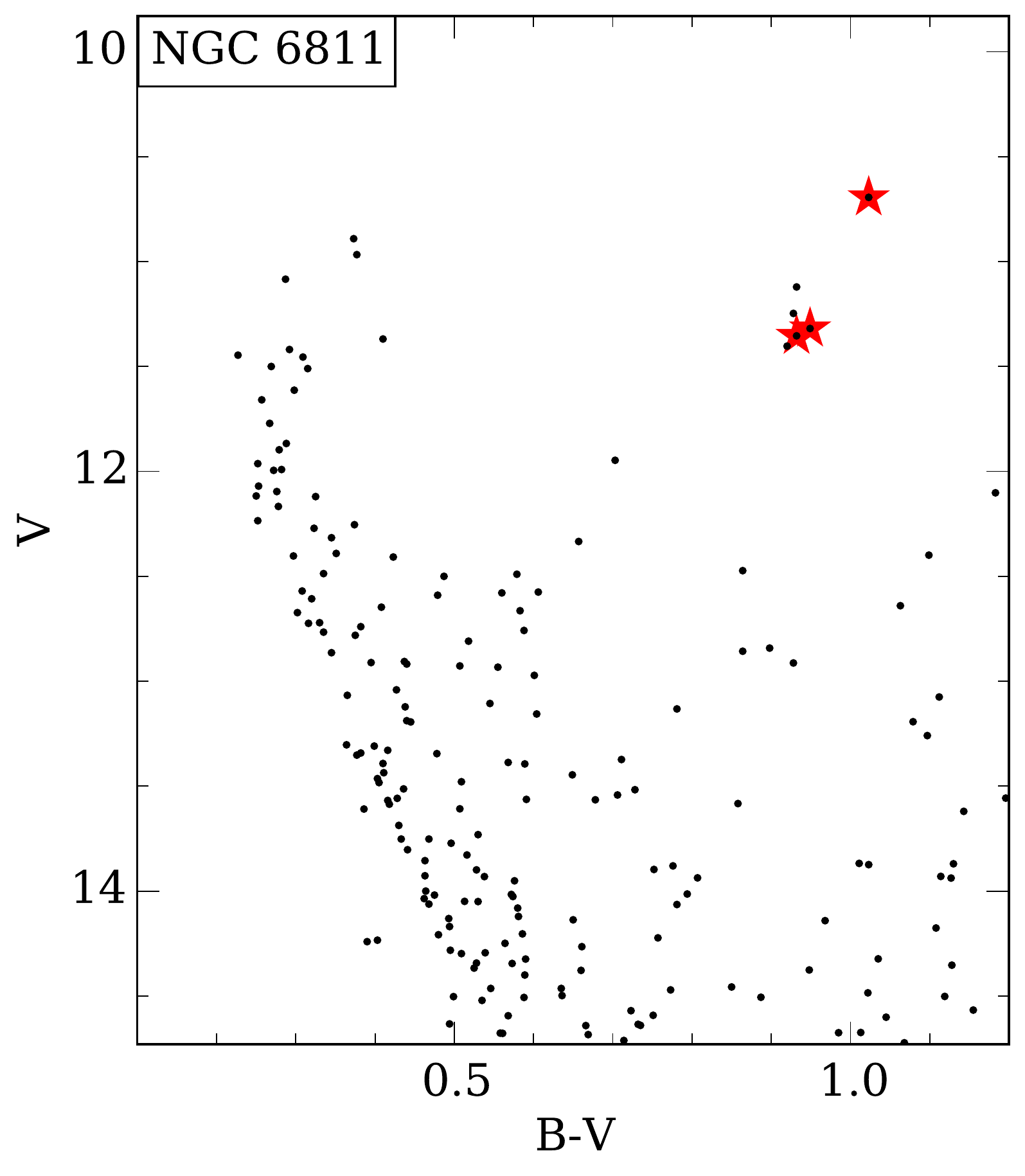}
\includegraphics[width=.24\textwidth]{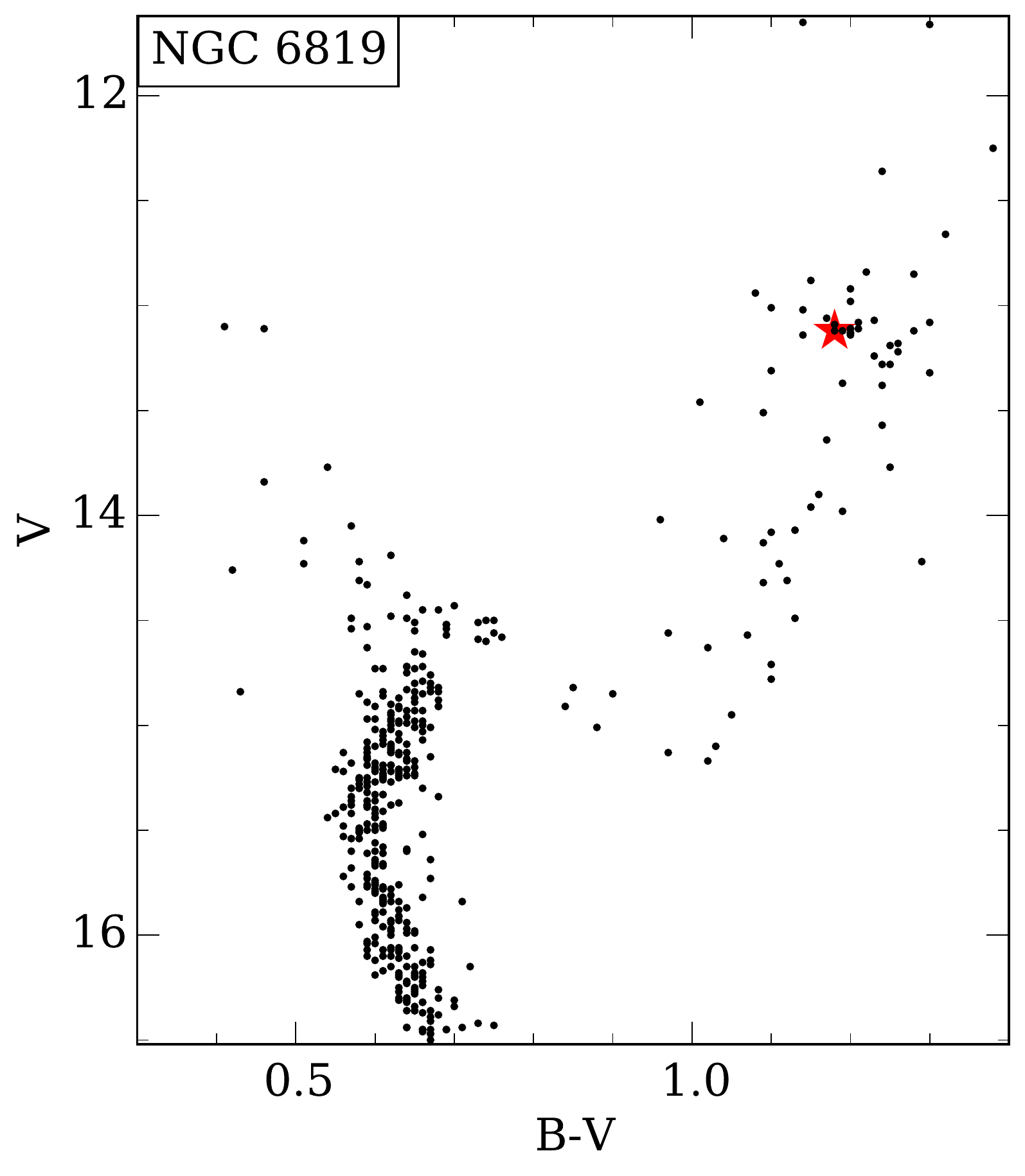}
\includegraphics[width=.24\textwidth]{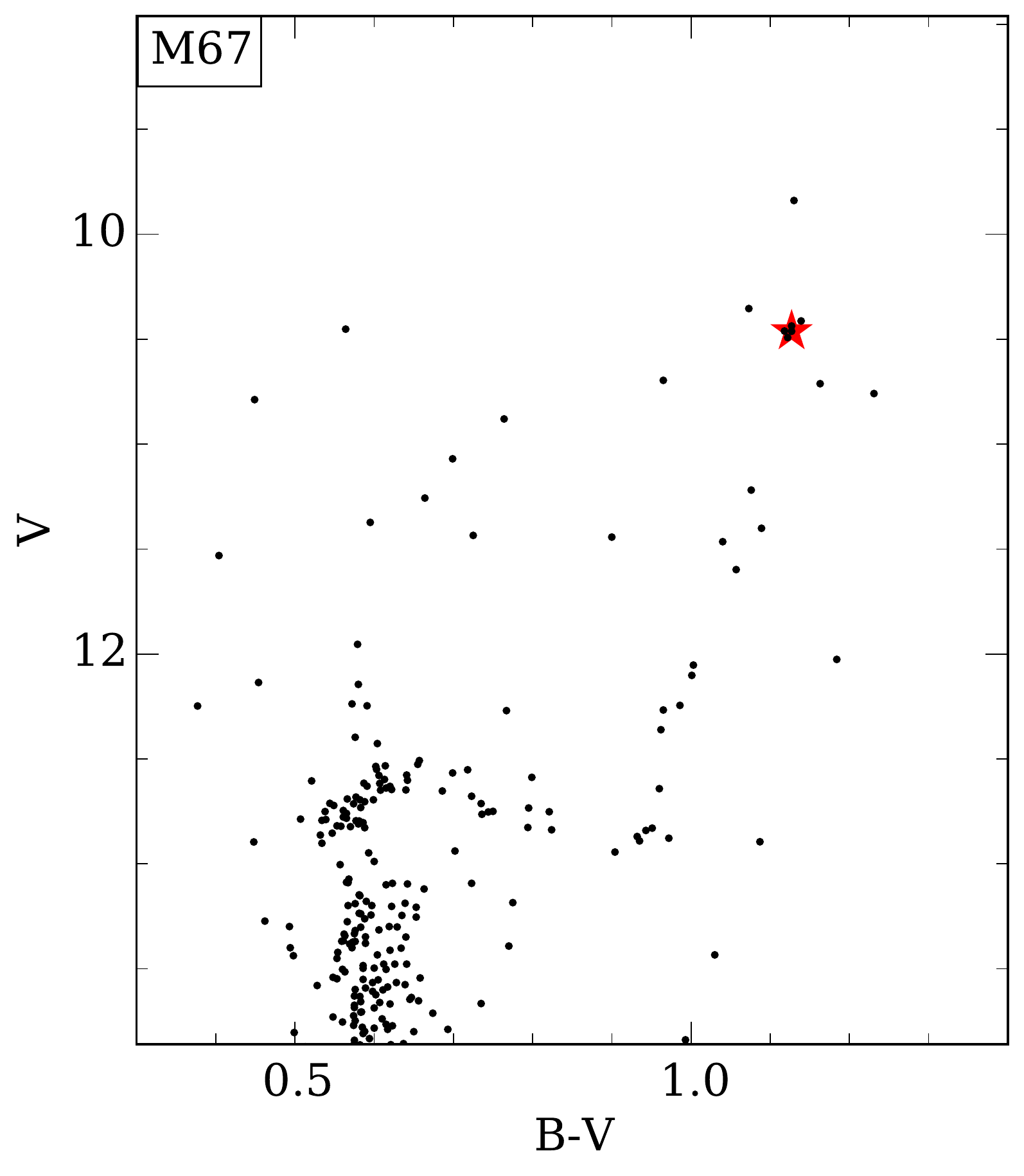}
\includegraphics[width=.24\textwidth]{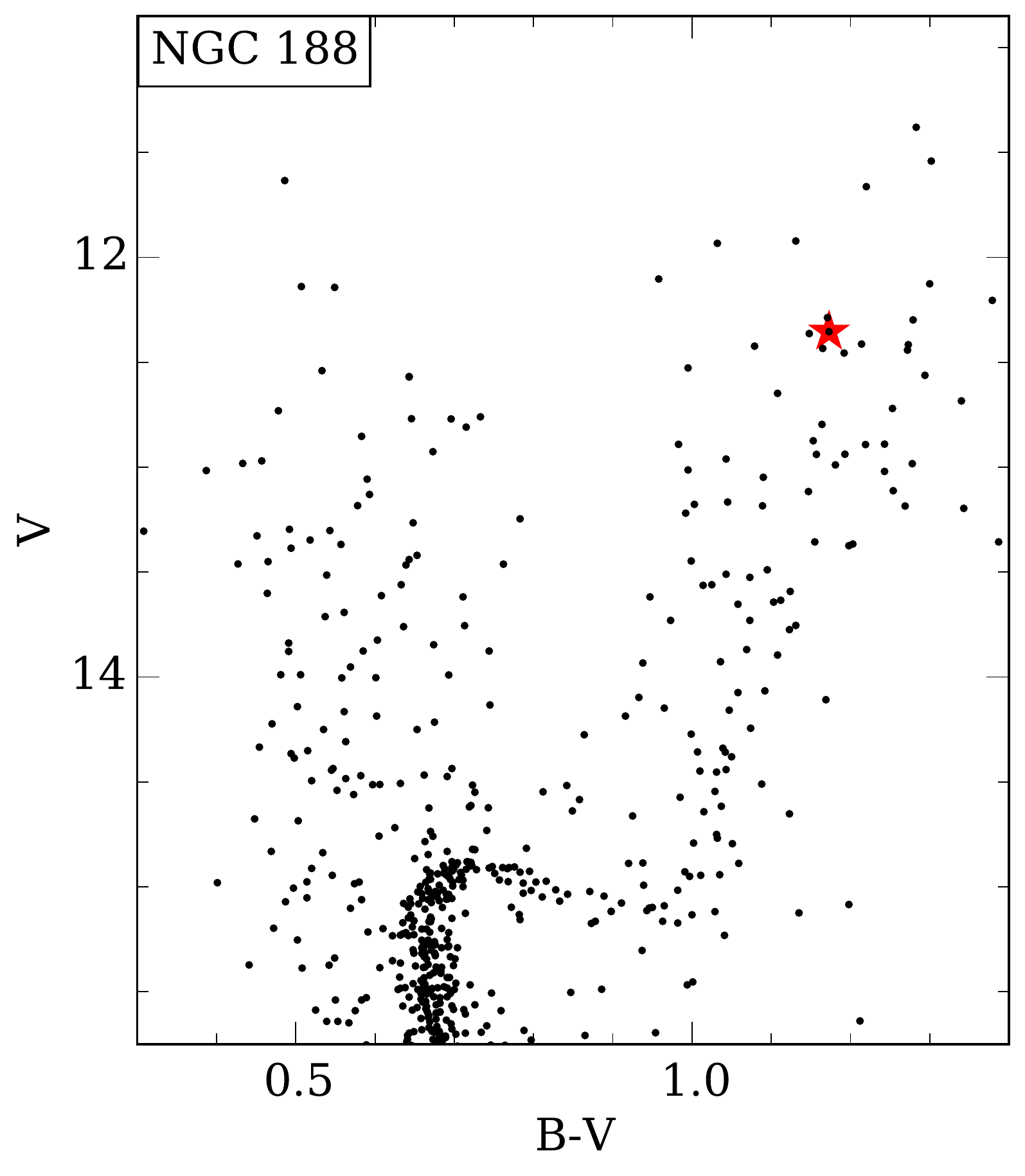}
\caption{From left to right: The CMDs for NGC\,6811, NGC\,6819, and M67 with photometry from \citet{Yontan2015, Hole2009, Yadav2008} respectively and lastly a CMD for NGC\,188 with photometry from Peter Stetson for one of his standard fields \citep{Stetson2004}. The target star(s) for each cluster is marked with a red star.}
\label{fig:cmd}
\end{figure*}

Parameters derived from the photometry presented in Fig.~\ref{fig:cmd} were used as starting estimates, listed in Table~\ref{tab:param_phot}. The effective temperatures for NGC\,6819, M67 and NGC\,188 were calculated from the color-temperature calibration presented by \citet{Ramirez2005} for different filter combinations. For each target, an average of the different filter combinations was used. For NGC\,6811 we used the effective temperatures from \citet{Arentoft2017} as first estimates.

\begin{table}\small
        \centering
        \caption{\textsl{Adopted parameters from photometry and asteroseismology}}
        \begin{threeparttable}
                \begin{tabular}{l c c c}  
                        \toprule
                        Target & $T_\text{eff}$ [K] & $\nu_\text{max}$ & $\log g$\\
                        \midrule
                        NGC\,6811-KIC9655167\tnote{a} & 4924 $\pm$ 100 & 99.4  $\pm$ 2.1 & 2.91 $\pm$ 0.01 \\
                        NGC\,6811-KIC9716090          & 4980 $\pm$ 100 & 107.8 $\pm$ 1.4 & 2.94 $\pm$ 0.01 \\
                        NGC\,6811-KIC9716522          & 4826 $\pm$ 100 & 53.7  $\pm$ 1.0 & 2.64 $\pm$ 0.01 \\  
                        NGC\,6819-KIC5024327          & 4698 $\pm$ 33  & 43.9  $\pm$ 1.8\tnote{b} & 2.55 $\pm$ 0.02 \\
                        M67-EPIC211415732             & 4718 $\pm$ 43  & 37.9  $\pm$ 2.6\tnote{c} & 2.48 $\pm$ 0.06 \\
                        NGC\,188-5085                 & 4637 $\pm$ 19  & -               & 2.45  $\pm$ 0.14 \\
                        \bottomrule
                \end{tabular}
                \footnotesize
                \begin{tablenotes}
                \item[a] All NGC\,6811 values are from \citet{Arentoft2017}
                \item[b] \citet{Corsaro2012}
                \item[c] \citet{Stello2016}
                \end{tablenotes}
        \end{threeparttable}
        \label{tab:param_phot}
\end{table}

The $\log g$ for the stars in NGC\,6811, NGC\,6819 and M67 in Table~\ref{tab:param_phot} is from asteroseismology, (\citealt{Arentoft2017,Corsaro2012,Stello2016} respectively) calculated with the $\nu_\text{max}$ scaling relations (Brown1991, Kjeldsen1995): 
\begin{align}
\log g = \log \left( \left( \frac{\nu_\text{max}}{3100} \right) \cdot \left( \frac{T_\text{eff}}{5777} \right) ^{1/2} \right) + 4.44 \ .
\label{eq:seis}
\end{align}
The asteroseismic $\log g$ values are determined to very high precision and have been shown to be in very close agreement with the physical $\log g$ (e.g., \citealt{Brogaard2016,Gaulme2016,Frandsen2013}). This provides a strong constraint on the analysis. For NGC\,188, the $\log g$ is calculated with the equation from \citet{Nissen1997} using a mass of 1.1$M_\odot$ from \citet{Meibom2009} assuming no significant mass loss on the RGB \citep{Miglio2012}.

We have tested different line lists and different programs to calculate the equivalent widths, which will be discussed in more detail in a forthcoming paper. Based on external constraints from especially asteroseismology, the final choice of line list is from \citet{Carraro2014a} with astrophysical $\log gf$ values based on solar abundances from \citet{Grevesse1998}. We have omitted lines stronger than \SI{100}{m\AA} for Fe and \SI{120}{m\AA} for other elements. A few additional magnesium and yttrium lines were added to do a more robust determination of [Y/Mg] for the [Y/Mg] vs. age relation discussed in Sect.~\ref{sec:ymgage}. The final line list will be given in a forthcoming paper with the measured equivalent widths for all lines. The equivalent widths were measured with DAOSPEC \citep{Stetson2008} and the auxiliary program Abundance with SPECTRUM \citep{Gray1994} was used to calculate the atmospheric parameters and abundances based on solar abundances from \citet{Grevesse1998} and ATLAS9 stellar atmosphere models \citep{Castelli2004}. Local thermodynamic equilibrium (LTE) is assumed. There may be non-LTE effects on the derived abundances, but since the stars have similar parameters, differential abundances between the stars are reliable.

The atmospheric parameters were determined by requiring that [Fe/H] has no systematic dependence on the excitation potential or the strength of the FeI lines and that the mean [Fe/H] values derived from FeI and FeII lines are consistent. The slope of [Fe/H] as a function of excitation potential is sensitive to the effective temperature and the slope of [Fe/H] as a function of the reduced equivalent widths of the lines ($\log (\text{EW})/\lambda$) depends on the microturbulence. The surface gravity is determined via its effect on the electron pressure in the stellar atmosphere with the FeI and FeII equilibrium, as the FeII lines are more sensitive to pressure changes than the FeI lines. This is however also affected by the temperature and heavier element abundances and it was necessary to make a number of iterations. For NGC\,6811, NGC\,6819 and M67, we also calculated a new asteroseismic $\log g$ with the newly found effective temperature, but the variation in effective temperature is low enough, that the asteroseismic $\log g$ was not significantly affected.

\section{Atmospheric parameters and abundances}
\label{sec:results}

The final result for the atmospheric parameters are presented in Table~\ref{tab:param_spec}. The uncertainties are only internal and calculated by varying a parameter until at least a 3$\sigma$ uncertainty is produced on either of the two slopes, [Fe/H] vs. excitation potential and [Fe/H] vs. reduced equivalent width, or on the difference between FeI and FeII. The change in the parameter is then divided by the highest produced uncertainty to give one standard deviation, provided in Table~\ref{tab:param_spec}. The errorbars on [Fe/H], [$\alpha$/Fe] and [Y/Mg] is the standard error of the mean. 

\begin{figure*}
\centering
\includegraphics[width=.45\textwidth]{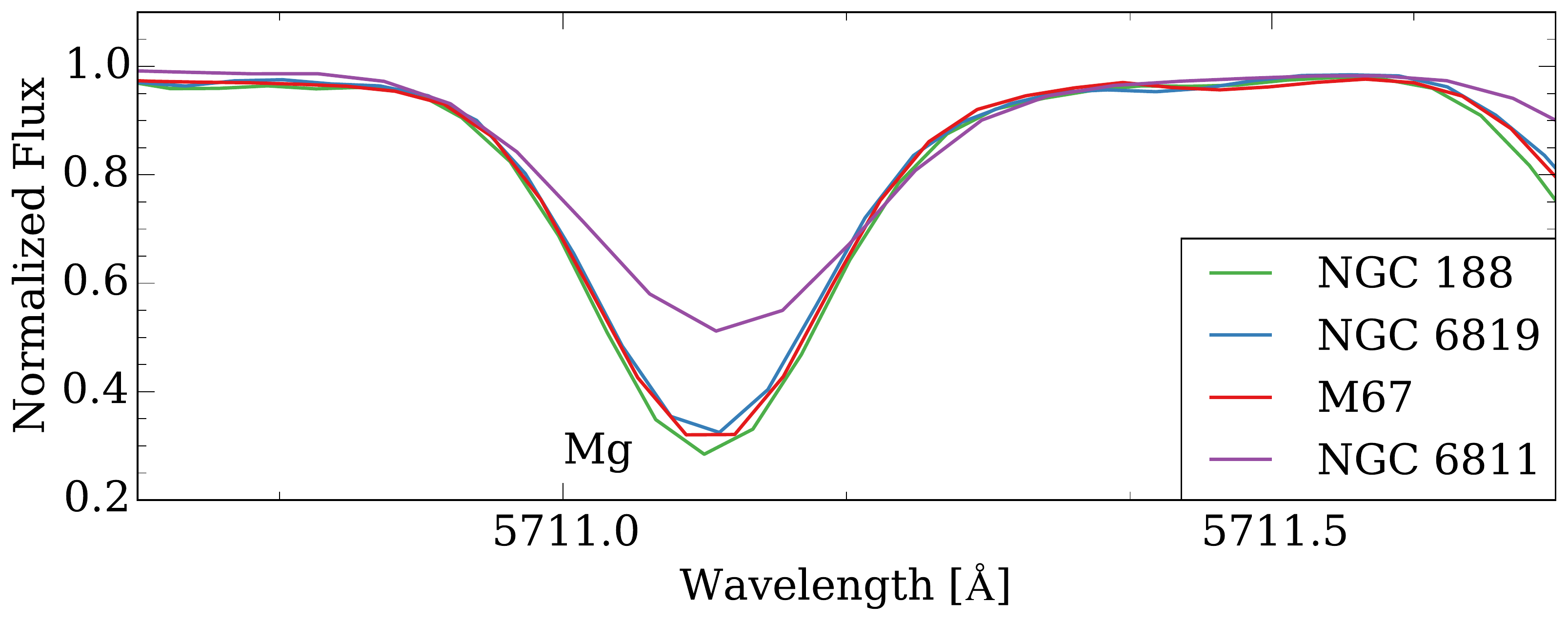}
\includegraphics[width=.45\textwidth]{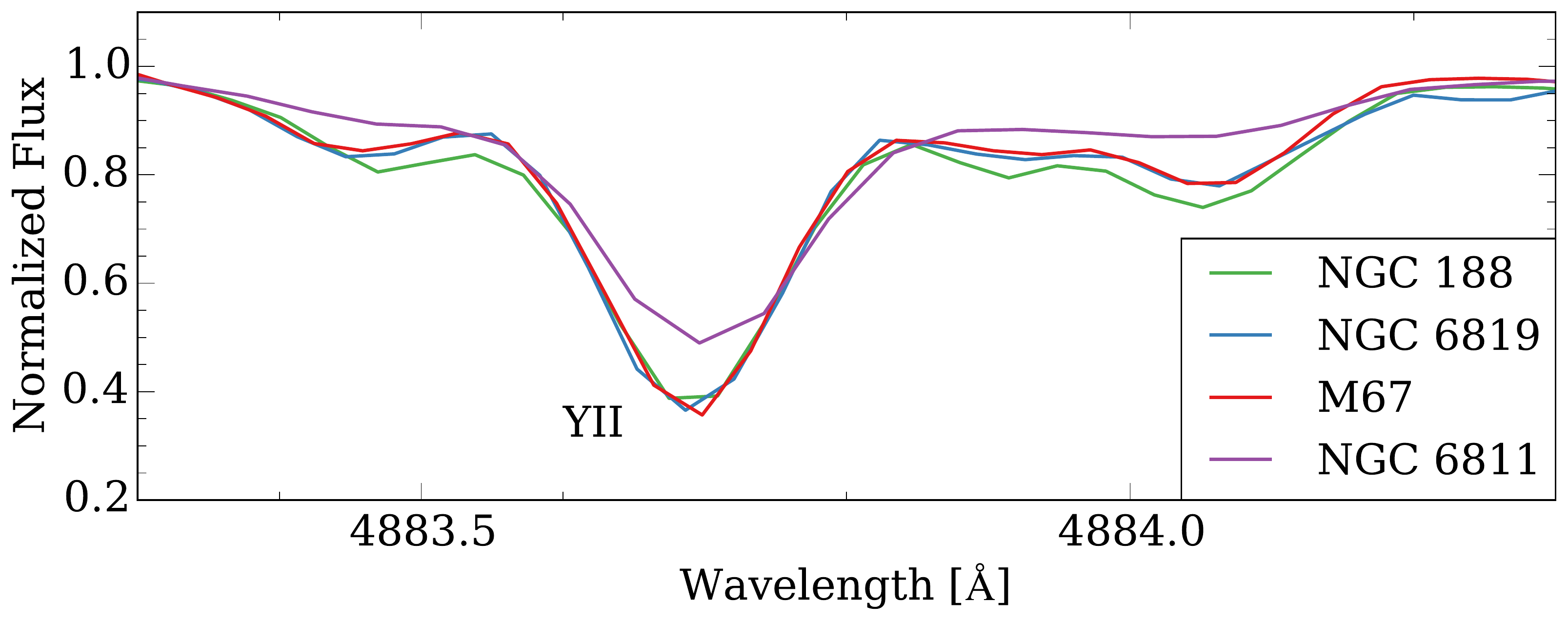}
\caption{Two parts of the normalized spectra for one target in each cluster to illustrate the high quality of the spectra. NGC\,6819, M67 and NGC\,188 are very similar, which is also illustrated in the atmospheric parameters in Table~\ref{tab:param_spec}. NGC\,6811 spectrum is for KIC9655167.}
\label{fig:spec}
\end{figure*}

The spectroscopic $\log g$ values are, within the errorbars, in agreement with the results from asteroseismology. The metallicities for all targets are close to solar, with NGC\,188 having a slightly higher-than-solar abundance and NGC\,6811 having a slightly lower-than-solar abundance, which along with the temperature differences can be seen in the line depths in Fig.~\ref{fig:spec}.  

The metallicity of NGC\,6819 is lower than that found by \citet{Bragaglia2001} ([Fe/H]=$+0.09 \pm 0.03$ dex found by analyzing three giants with high resolution spectroscopy), but fits better with the value from \citet{Lee-Brown2015} who performed an analysis of multiple targets but with low resolution spectroscopy. Their value of [Fe/H]= $-0.02 \pm 0.02$ dex is found using main sequence and turnoff stars.

\begin{table*}\small
        \centering
        \caption{\textsl{Atmospheric parameters and abundances from spectroscopy }}
        \begin{threeparttable}
                \begin{tabular}{l c c c c c c c}  
                        \toprule
                        Target & $T_\text{eff}$ [K] & $\log g$ & $v_t$ [km/s] & $[$Fe/H] & $n$\tnote{a} & $[\alpha$/Fe] & $[$Y/Mg] \\
                        \midrule
                        NGC\,6811-KIC9655167 & 5085 $\pm$ 22 & 2.96 $\pm$ 0.08 & 1.09 $\pm$ 0.05 & -0.08 $\pm$ 0.01 & 87/11 & 0.02 $\pm$ 0.02 & 0.09 $\pm$ 0.05\\
                        NGC\,6811-KIC9716090 & 5115 $\pm$ 25 & 2.93 $\pm$ 0.08 & 1.03 $\pm$ 0.07 & -0.11 $\pm$ 0.01 & 89/11 & 0.06 $\pm$ 0.03 & 0.23 $\pm$ 0.06\\
                        NGC\,6811-KIC9716522 & 4880 $\pm$ 30 & 2.60 $\pm$ 0.10 & 1.27 $\pm$ 0.06 & -0.12 $\pm$ 0.01 & 87/11 & 0.03 $\pm$ 0.02 & 0.12 $\pm$ 0.05\\
                        NGC\,6819-KIC5024327 & 4695 $\pm$ 18 & 2.52 $\pm$ 0.05 & 1.19 $\pm$ 0.05 & -0.02 $\pm$ 0.01 & 97/12 & -0.02 $\pm$ 0.02 & 0.08 $\pm$ 0.03\\
                        M67-EPIC211415732  & 4680 $\pm$ 18 & 2.43 $\pm$ 0.04 & 1.21 $\pm$ 0.05 & -0.03 $\pm$ 0.01 & 97/12 & 0.03 $\pm$ 0.02 & 0.01 $\pm$ 0.03\\
                        NGC\,188-5085        & 4580 $\pm$ 23 & 2.51 $\pm$ 0.06 & 1.17 $\pm$ 0.06 & 0.04 $\pm$ 0.01 & 95/12 & 0.00 $\pm$ 0.02 & -0.06 $\pm$ 0.06\\
                        \bottomrule
                \end{tabular}
                \footnotesize
                \begin{tablenotes}
                        \item[a] The number of FeI/FeII lines used.
                \end{tablenotes}
        \end{threeparttable}
        \label{tab:param_spec}
\end{table*}

The abundances for [Fe/H], $[\alpha$/Fe], and [Y/Mg] are given in Table~\ref{tab:param_spec} while several additional elements will be presented in a forthcoming paper.
The magnesium lines used are at wavelengths \SI{5711.09}{\AA}, \SI{6318.70}{\AA}, \SI{6319.23}{\AA} and \SI{6319.48}{\AA}. The yttrium lines are at \SI{4883.70}{\AA}, \SI{4900.13}{\AA} and \SI{5728.89}{\AA}. Not all of the lines were usable for each star. The magnesium line at \SI{5711.09}{\AA} (see Fig.~\ref{fig:spec}) is, for some of the targets, stronger than the limit of \SI{120}{m\AA} for non-iron lines, but it is well isolated for all stars, and we therefore chose to include it to do a more robust determination of Magnesium. The final value for [Y/Mg] used for NGC\,6811 is a the mean of the three targets.

\section{[Y/Mg] vs. age}
\label{sec:ymgage}

\begin{figure}
\centering
\includegraphics[width=.9\columnwidth]{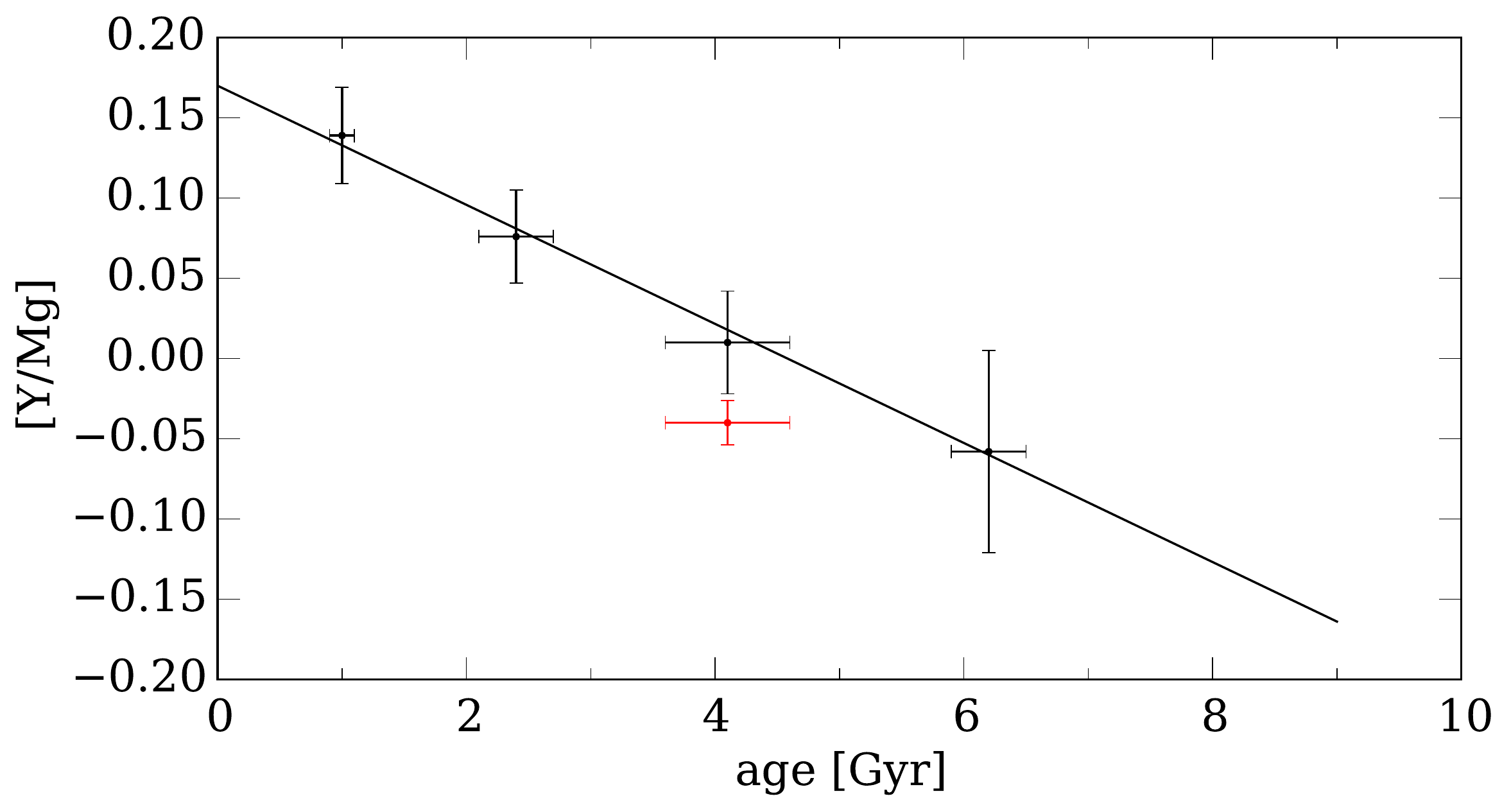}
\caption{[Y/Mg] vs. age for the four clusters. The line is the relation found by \citet{Nissen2016} in Eq.~\ref{eq:ygmage} for solar twins. The red point is an average of the results from \citet{Onehag2014} and the errorbar is the standard error of the mean for their 14 stars.}
\label{fig:ymgage}
\end{figure}

Magnesium is an alpha-element mostly originating from type II supernovae explosions, which gives an increase of [Mg/Fe] with increasing stellar age because iron is also produced in the later type Ia supernovae explosions. Yttrium is an s-process element and [Y/Fe] is observed to decrease with increasing stellar age. This is likely a consequence of intermediate mass asymptotic giant branch stars not yet being important for the production of Y at early times. The slope of [Y/Fe] with age is steep and opposite to that of [Mg/Fe]. For solar twin stars, \citet{Nissen2016} found the relation:
\begin{align}
[\text{Y/Mg}] = 0.170(\pm0.009) - 0.0371(\pm 0.0013) \cdot \text{Age [Gyr].}
\label{eq:ygmage}
\end{align}
This relation is plotted in Fig.~\ref{fig:ymgage}. \citet{Feltzing2016} confirmed the relation for dwarfs of solar metallicity but found that it disappears for stars with [Fe/H]$\approx -0.5$ and below. We have extended the parameter range to helium-core-burning giants at close to solar metallicity, and they also follow the relation from \citet{Nissen2016} as seen in Fig.~\ref{fig:ymgage}. 
This result is of particular interest for galactic archaeology studies as giants are much brighter than dwarfs, which allows us to study farther regions of the galaxy and not only the solar neighborhood.  

\citet{Onehag2014} carried out an abundance analysis of 14 stars in M67 at different evolutionary stages with high resolution spectra. They find an average metallicity of [Fe/H]=+0.06 with an average [Y/Mg]=$-0.04 \pm 0.05$, which is a little lower than our result, but still close to being in agreement with the relation from \citet{Nissen2015}, marked in Fig.~\ref{fig:ymgage}

\section{Conclusion}
\label{sec:conclusion}

Atmospheric parameters and abundances have been determined for the four open clusters, NGC\,6811, NGC\,6819, M67 and NGC\,188 with an equivalent width analysis of individual spectral lines from high-resolution, high S/N observations from the NOT and the Keck I Telescope. The parameters obtained fit very well with the literature, and especially, the $\log g$ values fits with predictions from asteroseismology. The metallicities of all four clusters are nearly solar, with NGC\,6811 being slightly sub-solar.

The empirical relation between [Y/Mg] and age as presented by \citet{Nissen2016} was found to hold also for helium-core-burning giants of close to solar metallicity. This is of great importance to galactic chemical evolution studies, as the brighter nature of giants allows us to probe the Galaxy to greater distances and not only the solar neighborhood.

\begin{acknowledgements}
The authors wish to recognize and acknowledge the very significant cultural role and reverence that the summit of Mauna Kea has always had within the indigenous Hawaiian community.  We are most fortunate to have the opportunity to conduct observations from this mountain. Funding for the Stellar Astrophysics Centre is provided by The Danish National Research Foundation (Grant DNRF106). This research has made use of the SIMBAD database, operated at CDS, Strasbourg, France. MGP is funded from the European Research Council (ERC) under the European Union’s Horizon2020 research and innovation program (grant agreement N$^o$ 670519: MAMSIE).
\end{acknowledgements}

\bibliographystyle{aa}
\bibliography{/home/ditte/Dropbox/Speciale/library}

\end{document}